\documentclass[journal]{IEEEtran}
\usepackage{graphicx}
\usepackage{epsfig}
\usepackage{multirow}
\usepackage{subfigure}
\usepackage{amsmath,amssymb}
\usepackage[linesnumbered,ruled]{algorithm2e}
\usepackage{cite}
\usepackage{balance}
\usepackage{epsfig}
\usepackage{graphicx}
\usepackage{color}
\usepackage[english]{babel}
\usepackage{amsfonts,amsthm,amsmath}
\usepackage{times}
\usepackage{multirow}
\usepackage{subfigure}
\usepackage{amsmath}
\usepackage{amsfonts}
\usepackage{psfrag}

\usepackage[linesnumbered,ruled]{algorithm2e}

\theoremstyle{plain}

\newtheorem{remark}{Remark}

\begin{document}

\title{Rate-Distributed Spatial Filtering Based Noise Reduction in Wireless Acoustic Sensor Networks}

\author{Jie Zhang, Richard Heusdens, and Richard C. Hendriks \\

\thanks{
Manuscript received xxxxx xx, 2017; revised xxxxx xx, 2017; accepted
xxxxx xx, 2017. Date of publication xxxxx xx, 2017; date of current version
xxxxx xx, 2017. This work is supported by the China Scholarship Council (NO. 201506010331) and Circuits and Systems (CAS) Group, Delft University of Technology, Delft, The Netherlands. The associate editor coordinating the review of this manuscript and approving it for publication was Prof. ********.}
\thanks{The authors are with the Faculty of Electrical Engineering, Mathematics and
Computer Science, Delft University of Technology, 2628 CD Delft, The Netherlands (e-mail:
j.zhang-7@tudelft.nl, r.heusdens@tudelft.nl, r.c.hendriks@tudelft.nl).}
\thanks{Color versions of one or more of the figures in this paper are available online
at http://ieeexplore.ieee.org.}
\thanks{Digital Object Identifier: **************}
}
\markboth{Report: Rate-Distributed Spatial Filtering Based Noise Reduction in Wireless Acoustic Sensor Networks}%
{Shell \MakeLowercase{\textit{et al.}}: Bare Demo of IEEEtran.cls for IEEE Journals}

\maketitle
\begin{abstract}
In wireless acoustic sensor networks (WASNs), sensors typically have a limited energy budget as they are often battery driven. Energy efficiency is therefore essential to the design of algorithms in WASNs. One way to reduce energy costs is to only select the sensors which are most informative, a problem known as {\it sensor selection}. In this way, only sensors that significantly contribute to the task at hand will be involved. In this work, we consider a more general approach, which is based on rate-distributed spatial filtering. Together with the distance over which transmission takes place,  bit rate directly influences the energy consumption. We try to minimize the battery usage due to transmission, while constraining the noise reduction performance. This results in an efficient rate allocation strategy, which depends on the underlying signal statistics, as well as the distance from sensors to a fusion center (FC). Under the utilization of a linearly constrained minimum variance (LCMV) beamformer, the problem is derived as a semi-definite program. Furthermore, we show that rate allocation is more general than sensor selection, and sensor selection can be seen as a special case of the presented rate-allocation solution, e.g., the best microphone subset can be determined by thresholding the rates. Finally, numerical simulations for the application of estimating several target sources in a WASN demonstrate that the proposed method outperforms the microphone subset selection based approaches in the sense of energy usage, and we find that the sensors close to the FC and close to point sources  are allocated with higher rates.
\end{abstract}
\IEEEpeerreviewmaketitle

\section{Introduction}\label{sec:intro}
\IEEEPARstart{R}{ecently}, wireless acoustic sensor networks (WASNs) have attracted an increasing amount of interest~\cite{bertrand2011applications,zeng2014distributed,cherkassky2017blind}. Compared to conventional microphone arrays with a fixed configuration, WASNs have advantages in array-size limitation and scalability of the networks.
In a WASN, each sensor node is equipped with a single microphone or a small microphone array, and the nodes are spatially distributed across a specific environment. Due to the fact that the microphone nodes in a WASN can be placed anywhere, the sound field is sampled in a much larger area. It is possible that some of the nodes are close to the target source(s) and have higher signal-to-noise ratio (SNR), such that higher quality recordings can be obtained. In a WASN, the microphone nodes are connected to their neighboring nodes or a fusion center (FC) using wireless links, resulting in a distributed  or centralized framework, respectively. In this work, we will mainly focus on the centralized framework, where each  node samples and quantizes the microphone recordings, and transmits them to a remote FC. The tasks of interest, e.g., signal estimation, noise reduction or binaural cue preservation, are assumed to occur at the FC.

In WASNs, each sensor node is usually battery powered having  a  limited energy budget. Energy consumption is therefore significant to the design of algorithms. Generally, the energy usage within the context of WASNs can be linked to two processes: data transmission and data processing~\cite{akyildiz2002wireless,Yick20082292}. The data transmission occurs between the nodes and the FC, and data processing at the FC end. Usually, data exchange is more expensive than data processing in terms of energy usage.

In order to reduce the energy usage in WASNs, there are two techniques that can be employed: sensor selection~\cite{chepuri2015sparsity,zhang2017microphone,joshi2009sensor} and rate allocation~\cite{roy2009rate,amini2016impact,de2016generalized}. For sensor selection, the most informative subset of sensors is chosen by maximizing a performance criterion while constraining the cardinality of the selected subset, or by minimizing the cardinality while constraining the performance. In this way, the number of sensors contained in the selected subset can be much smaller than the total set of sensors, resulting in a sparse selection. Due to the fact that only the selected sensors need to transmit their recordings to the FC, sensor selection is an effective way to save the energy usage.

Compared to sensor selection, rate allocation allows for a more smooth operating curve as sensors are not selected to only operate at full rate or zero rate (when not selected), but at any possible rate. For rate allocation, the idea is to allocate higher rates to the more informative sensors while lower or zero rates are allocated to the others. There are many works on quantization for signal estimation in the context of wireless sensor networks, see~\cite{xiao2006power,cui2007estimation} and reference therein, typically under the assumption that the measurement noise across sensors is mutually uncorrelated. These models are not suitable for realistic audio applications, e.g., speech enhancement, where the noise is typically correlated across sensors because of the presence of directional interfering sources. In~\cite{lawin2011analysis,amini2016impact}, the effect of a bit-rate constraint was investigated for noise reduction in WASNs. In~\cite{roy2009rate}, rate-constrained collaborative noise reduction for wireless hearing aids (HAs) was studied from an information-theoretic standpoint, resulting in an information transmission strategy between two nodes. However, the approach proposed in~\cite{roy2009rate}  requires full binaural statistics which are difficult to estimate in a practical setting. In~\cite{de2016generalized}, a greedy quantization method was proposed for speech signal estimation based on a so-called signal utility, which indeed represents the importance of microphone recordings. However, it only decreases/increases one bit for a node at each iteration, resulting in low convergence speed.

The difference between sensor selection and rate allocation problems lies in binary versus more smooth decisions. Given a maximum bit rate, the sensor selection approaches choose a subset of sensors first, and the selected sensors then communicate with the FC using the maximum rate. That is, each sensor only makes a binary decision on the communication rate, i.e., zero or maximum rate. In contrast to sensor selection, rate allocation approaches can execute multiple decisions on the communication rate, i.e., any bit rate can be fractional from zero bit rate to the maximum bit rate. If a sensor is allocated zero bits, it will not be selected. Hence, in general, rate allocation approaches do not lead to a WASN that is as sparse as the one that is obtained by the sensor selection approaches, but they can better reduce energy consumption used for transmission. On the other hand,  sensor selection approaches could save more energy usage for data processing at the FC end, as typically less measurements are involved in computations.

In this work, we will only consider the energy usage for data transmission  and neglect the energy usage for other processes. The wireless transmission power is regarded as a function of the distance between sensor nodes and the FC and the rate (i.e., bit per sample) which is used to quantize the signals to be transmitted. We intend to reduce energy usage from the perspective of rate allocation for spatial filtering based noise reduction in WASNs. We minimize the total wireless transmission costs by constraining the performance of the output noise power. Under the utilization of a linearly constrained minimum variance (LCMV) beamformer, the problem is solved by convex optimization techniques. After the allocated bit rates are determined, each microphone node uniformly quantizes and transmits its recordings to the FC for the signal processing tasks at hand.

\subsection{Contributions}
The contributions of the paper can be summarized as follows. Firstly, we design a rate allocation strategy for rate-distributed LCMV (RD-LCMV) beamforming in WASNs by minimizing the energy usage and constraining the noise reduction performance. The original non-convex optimization problem is relaxed using convex relaxation techniques and reformulated as semi-definite programming. Based on  numerical results in simulated WASNs, we find that the microphone nodes that are close to the sources (including target sources and interferers) and the FC are more likely to be allocated with more bit rates, because they have more information on SNR and cost less energy, respectively.

Secondly, we extend the model-driven microphone subset selection approach for minimum variance distortionless response (MD-MVDR) beamformer from~\cite{zhang2017microphone} to the LCMV beamforming framework (referred as MD-LCMV). By doing so, we find the link between rate allocation and sensor selection problems, i.e., rate allocation is a generalization of sensor selection. In~\cite{zhang2017microphone}, the best microphone subset is chosen by minimizing the total transmission costs and constraining the noise reduction performance, where the transmission cost between each node and the FC is only considered as a function of distance. The selected microphone will communicate with the FC using the maximum bit rate. The energy model of the approach in the current paper is more general as compared to that in~\cite{zhang2017microphone}. Based on the rates obtained by the proposed RD-LCMV approach, the best microphone subset of MD-LCMV can be determined by putting a threshold on the rates, e.g., the sensors whose rates are larger than this threshold are chosen.

Finally, numerical simulations demonstrate that the selected microphone subsets resulting from thresholding the rates from the RD-LCMV method and directly applying  MD-LCMV are completely the same. Both RD-LCMV and MD-LCMV can guarantee a given performance requirement, but RD-LCMV shows a superiority in energy efficiency.

\subsection{Outline and notation}
The rest of this paper is organised as follows. Sec.~\ref{sec:preliminary} presents preliminary knowledge on the signal model, uniform quantization, the used energy model and LCMV beamforming. In Sec.~\ref{sec:rateLCMV}, the problem formulation and a solver for the RD-LCMV optimization are given. Sec.~\ref{sec:rateLCMV_vs_SenSel} extends the sensor selection for MVDR beamforming from~\cite{zhang2017microphone} to the LCMV beamforming framework and discusses the link between sensor selection and rate allocation problems. Sec.~\ref{sec:exp_example} shows the application of the proposed RD-LCMV method to the WASNs. Finally, Sec.~\ref{sec:conclusion} concludes this work.

The notation used in this paper is as follows: Upper (lower) bold face letters are used for matrices (column vectors). $(\cdot)^T$ or $(\cdot)^H$ denotes (vector/matrix) transposition or conjugate transposition. ${\rm diag}(\cdot)$ refers to a block diagonal matrix with the elements in its argument on the main diagonal. $\mathbf{1}_N$ and $\mathbf{O}_N$ denote the $N\times 1$ vector of ones and the $N\times N$ matrix with all its elements equal to zero, respectively. $\mathbf{I}_N$ is an identity matrix of size $N$. $\mathbb{E}\{\cdot\}$ denotes the statistical expectation operation. $\mathbf{A}\succeq \mathbf{B}$ means that $\mathbf{A}-\mathbf{B}$ is a positive semidefinite matrix. Finally, $\odot$ denotes the Hadamard (elementwise) product.

\section{Preliminaries}\label{sec:preliminary}
In this section, we introduce some preliminary concepts related to rate-distributed spatial filtering in WASNs.
\subsection{Signal model}
We consider a spatially distributed candidate set of $M$ microphone sensors that collect, quantize and transmit their observations to an FC. In the short-term Fourier transform (STFT) domain, let $l$ denote the frame index and $\omega$ the frequency bin index, respectively. We assume that the user (i.e., FC) has $\mathcal{I}$ speech sources of interest, while $\mathcal{J}$ interfering sources are potentially present in the environment. Using an STFT-domain description, the noisy DFT coefficient of the quantized signal which is to be transmitted to the FC at the $k$th microphone, say $\hat{y}_k(\omega, l), k = 1,2, \cdots ,M$, is given by
\begin{equation} \label{eq:signal_model_scale}
 \hat{y}_k(\omega,l) = y_k(\omega,l) + q_k(\omega,l), \forall k,
\end{equation}
where $q_k(\omega,l)$ denotes the quantization noise which is assumed to be uncorrelated with the microphone recording\footnote{This assumption holds under high rate communication. Under low rate, this can be achieved using subtractive dither~\cite{amini2016impact,gray1993dithered}.} $y_k(\omega,l)$\footnote{In real-life applications, $y_k$ is already quantized, since it is acquired by the analog-to-digital converter (ADC) of the $k$th microphone. In this case, $q_k$ would represent the error from changing the bit resolution of $y_k$. }, given by
\begin{equation}
y_k(\omega,l) =\sum_{i=1}^{\mathcal{I}} \underbrace{a_{ik}(\omega)s_i(\omega,l)}_{x_{ik}(\omega,l)}+ \sum_{j=1}^{\mathcal{J}}\underbrace{b_{jk}(\omega)u_j(\omega,l)}_{n_{jk}(\omega,l)} + v_k(\omega,l),
\end{equation}
with
\begin{itemize}
\item $a_{ik}(\omega)$ denoting the acoustic transfer function (ATF) of the $i$th target signal with respect to the $k$th microphone;
\item $s_i(\omega,l)$ and $x_{ik}(\omega,l)$, the $i$th target source at the source location and the
$i$th target source at the $k$th microphone, respectively;
\item $b_{jk}(\omega)$ the ATF of the $j$th interfering source with respect to the $k$th microphone;
\item $u_j(\omega,l)$ and $n_{ik}(\omega,l)$, the $j$th interfering source at the source location and the
$j$th interference source at the $k$th microphone, respectively;
\item $v_k(\omega,l)$ the $k$th microphone self noise.
\end{itemize}
For notational convenience, we will omit the frequency variable $\omega$ and the frame index $l$  now onwards bearing in mind that the processing takes place in the STFT domain.
Using vector notation, the $M$ channel signals are stacked in a vector $\hat{\mathbf{y}} = [\hat{y}_1,...,\hat{y}_M]^T  \in \mathbb{C}^M$.
Similarly, we define $M$  dimensional vectors $\mathbf{y}, \mathbf{x}_i,\mathbf{n}_j, \mathbf{v},\mathbf{q}$ for the microphone recordings, the $i$th target component, the $j$th interfering component, the additive noise and the quantization noise, respectively, such that the signal model in (\ref{eq:signal_model_scale}) can compactly be written as
\begin{equation}\label{eq:signal_model_vector}
 \hat{\mathbf{y}} = \mathbf{y} + \mathbf{q}= \sum_{i=1}^{\mathcal{I}}\mathbf{x}_i + \sum_{j=1}^{\mathcal{J}}\mathbf{n}_j + \mathbf{v} + \mathbf{q},
\end{equation}
where $\mathbf{x}_i = \mathbf{a}_i s_i \in \mathbb{C}^M$ with $\mathbf{a}_i=[a_{i1},a_{i2},\cdots,a_{iM}]^T$ and $\mathbf{n}_j = \mathbf{b}_j u_j \in \mathbb{C}^M$ with $\mathbf{b}_j=[b_{j1},b_{j2},\cdots,b_{jM}]^T$.
Alternatively, if we stack the ATFs for the target sources and the interfering sources, in matrices, the microphone recordings can also be written like,
\begin{equation}\label{eq:signal_model_matrix}
\mathbf{y}= \mathbf{A}\mathbf{s} + \mathbf{B}\mathbf{u} + \mathbf{v},
\end{equation}
where $\mathbf{A}=[\mathbf{a}_1,\cdots,\mathbf{a}_{\mathcal{I}}]\in \mathbb{C}^{M\times \mathcal{I}}, \mathbf{s}=[s_1,\cdots,s_{\mathcal{I}}]^T\in \mathbb{C}^{\mathcal{I}}, \mathbf{B} = [\mathbf{b}_1,\cdots,\mathbf{b}_{\mathcal{J}}]\in \mathbb{C}^{M\times \mathcal{J}}, \mathbf{u}=[u_1,\cdots,u_{\mathcal{J}}]^T \in \mathbb{C}^{\mathcal{J}}$. In order to focus on the concept of rate-distributed noise reduction, we assume in this work  that the ATFs  of the existing sources (i.e., $\mathbf{A}$ and $\mathbf{B}$) are known.

Assuming that the target signals and the interferers are mutually uncorrelated, the correlation matrix of the recorded signals is given by
\begin{equation}\label{eq:Ryy=Rxx+Rnn}
\vspace{-0.1cm}
  \mathbf{R}_{\mathbf{y}\mathbf{y}} = \mathbb{E}\{\mathbf{y}\mathbf{y}^H\} = \mathbf{R_{xx}} + \underbrace{\mathbf{R_{uu}}+ \mathbf{R_{vv}}}_{\mathbf{R_{nn}}} \in \mathbb{C}^{M\times M},
  \vspace{-0.05cm}
\end{equation}
where  $\mathbf{R_{xx}}=\sum_{i=1}^{\mathcal{I}}\mathbb{E}\{\mathbf{x}_i\mathbf{x}_i^H\}=\sum_{i=1}^{\mathcal{I}}P_{s_i}\mathbf{a}_i\mathbf{a}_i^H
=\mathbf{A}\mathbf{\Sigma}_{\mathbf{x}}\mathbf{A}^H$ with $P_{s_i}=\mathbb{E}\{|s_i|^2\}$ representing the power spectral density (PSD) of the $i$th target source and $\mathbf{\Sigma}_{\mathbf{x}}={\rm diag}\left([P_{s_1},\cdots,P_{s_\mathcal{I}}] \right)$. Similarly, $\mathbf{R_{uu}}=\sum_{j=1}^{\mathcal{J}}\mathbb{E}\{\mathbf{n}_i\mathbf{n}_i^H\}=\sum_{j=1}^{\mathcal{J}}P_{u_i}\mathbf{b}_j\mathbf{b}_j^H
=\mathbf{B}\mathbf{\Sigma}_{\mathbf{u}}\mathbf{B}^H$ with $P_{u_i}=\mathbb{E}\{|u_i|^2\}$ representing the PSD of the $j$th interfering source and $\mathbf{\Sigma}_{\mathbf{u}}={\rm diag}\left([P_{u_1},\cdots,P_{u_\mathcal{J}}] \right)$.
In practice, $\mathbf{R_{nn}}$ can be estimated using sufficient noise-only segments, and $\mathbf{R_{xx}} = \mathbf{R_{yy}}-\mathbf{R_{nn}}$ can be estimated using the speech+noise segments\footnote{The statistics $\mathbf{R_{nn}}$ should be estimated using the noise-only segments under high-rate communication between sensors and the FC to guarantee its accuracy. However, the estimation of $\mathbf{R_{xx}}$ is not dependent on the communication rate, because it is obtained by subtracting $\mathbf{R_{nn}}$ from $\mathbf{R_{yy}}$ and both $\mathbf{R_{nn}}$ and $\mathbf{R_{yy}}$ have quantization noise included.}. The correlation matrix of all disturbances including quantization noise in the quantized signals $\hat{\mathbf{y}}$ is given by
\begin{equation}
\vspace{-0.1cm}
\mathbf{R_{n+q}} = \mathbf{R_{nn}} + \mathbf{R_{qq}},
\end{equation}
under the assumption that the received noises and quantization noise are mutually uncorrelated.
%
\vspace{-0.2cm}
\subsection{Uniform quantization}\label{sec:quantize}
The uniform quantization of a real number $a\in [-\frac{\mathcal{A}_k}{2}, \frac{\mathcal{A}_k}{2}]$ with $\mathcal{A}_k$ denoting the maximum absolute value of the $k$th microphone signal using $b_k$ bits can be expressed as
\begin{equation}
\vspace{-0.1cm}
Q(a) = \Delta_k \left(\left\lfloor\frac{a}{\Delta_k} \right\rfloor + \frac{1}{2} \right), \quad k=1,\cdots,M,
\end{equation}
where the uniform intervals have width $\Delta_k = \mathcal{A}_k/2^{b_k}$. Note that $\mathcal{A}_k$ is different from sensor to sensor which is determined by its own signal observations. Each sensor should inform its $\mathcal{A}_k$ to the FC by communication.
Considering the case of uniform quantization, the variance or PSD of the quantization noise is approximately given by~\cite{sripad1977necessary,gray1990quantization}
\begin{equation}
\vspace{-0.1cm}
\sigma_{q_k}^2 = \Delta_k^2/12, \quad k=1,\cdots,M,
\end{equation}
and the correlation matrix of the quantization noise across microphones  reads
\begin{equation}\label{eq:corr_matrix_quanti}
\vspace{-0.1cm}
\mathbf{R}_{\mathbf{qq}} = \frac{1}{12}\times {\rm diag}\left( \left[\frac{\mathcal{A}_1^2}{4^{b_1}}, \frac{\mathcal{A}_2^2}{4^{b_2}},...,\frac{\mathcal{A}_M^2}{4^{b_M}} \right] \right).
\vspace{-0.1cm}
\end{equation}
\vspace{-0.25cm}
\subsection{Transmission energy model}
We assume that the noise on the communication channels between the sensors and the FC is additive and white Gaussian with PSD $V_k$. The channel power attenuation factor is $d_k^r$, where $d_k$ is the transmission distance from the $i$th microphone to the FC and $r$ is the path loss exponent (typically $2\leq r\leq 6$)~\cite{shah2013adaptive,li2002detection}. Without loss of generality, we assume $r=2$ in this work. The SNR\footnote{The SNR mentioned in this section is used to measure the noise level over the communication channels, which is different from the acoustic noise or acoustic SNR that will be discussed in the experiments.} of the $k$th channel then is
\begin{equation}
 {\rm SNR}_k = d_k^{-2}E_k/V_k,
\end{equation}
where $E_k$ represents the transmitted energy of the $k$th microphone node per time-frequency sample.
Assuming Gaussian distributions for the noise and transmitted signal, the  maximum capacity of such a communication channel for a specific time-frequency bin is given by the Shannon theory~\cite{shannon1949communication}
\begin{equation}\label{eq:shannon-theory}
 b_k = \frac{1}{2}\log_2\left(1+{\rm SNR}_k \right),
\end{equation}
which implies that $b_k$ bits per sample at most can reliably be transmitted from microphone $k$ to the FC.
Based on the ${\rm SNR}_k$ and $b_k$, the transmission energy from microphone $k$ to the FC for a specific time-frequency bin  can be formulated as
\begin{equation}
E_k = d_k^2 V_k (4^{b_k}-1),
\end{equation}
which is a commonly used transmission model~\cite{shah2013adaptive,huang2007multihop,huang2009energy}.
The above transmission energy model holds under two conditions~\cite{shah2013adaptive,huang2009energy}: 1) in the context of spectrum-limited applications (e.g., audio signal processing); 2) under the assumption  that we quantize the microphone recordings at the channel capacity, which is in fact an upper bound.

\subsection{LCMV beamforming}
The well-known LCMV beamformer is a typical spatial filtering technique where the output noise energy is minimized under a set of linear constraints. These constraints can be used to preserve target sources, or steer zeros in the direction of interferences. In the context of binaural noise reduction~\cite{koutrouvelis2017relaxed,hadad2016binaural,hadad2015theoretical}, LCMV beamforming can be used to preserve certain interaural relations in order to preserve spatial cues.  Mathematically, the LCMV beamformer can be formulated as
\begin{equation}\label{eq:lcmv_opti}
\begin{aligned}
 \hat{\mathbf{w}}_{\rm LCMV}=\arg\min_{\mathbf{w}}\ \mathbf{w}^H\mathbf{R}_{\mathbf{n+q}}\mathbf{w}, \quad {\rm s.t. } \quad \mathbf{\Lambda}^H\mathbf{w}=\mathbf{f},
\end{aligned}
\end{equation}
which has $\mathcal{U}$ equality constraints with $\mathbf{f}=[f_1,f_2,\cdots,f_{\mathcal{U}}]^T\in \mathbb{C}^{\mathcal{U}}$  and $\mathbf{\Lambda} \in \mathbb{C}^{\mathcal{U}\times M}$. The closed-form solution to (\ref{eq:lcmv_opti}), which can be found by applying Lagrange multipliers, is given by~\cite{frost1972algorithm,van1988beamforming}
\begin{equation}\label{eq:lcmv_formula}
  \hat{\mathbf{w}}_{\rm LCMV} = \mathbf{R}_{\mathbf{n+q}}^{-1}\mathbf{\Lambda}\left(\mathbf{\Lambda}^H \mathbf{R}_{\mathbf{n+q}}^{-1}\mathbf{\Lambda}\right)^{-1}\mathbf{f}.
\end{equation}
The output noise power after LCMV beamforming can be shown to be given by~\cite{souden2010study}
\begin{equation}\label{eq:out_npower_lcmv}
\begin{aligned}
  \hat{\mathbf{w}}^H \mathbf{R}_{\mathbf{n+q}}\hat{\mathbf{w}} = \mathbf{f}^H \left(\mathbf{A}^H \mathbf{R}_{\mathbf{n+q}}^{-1}\mathbf{A}\right)^{-1} \mathbf{f} .
\end{aligned}
\end{equation}

\section{Rate-Distributed LCMV Beamforming}\label{sec:rateLCMV}
\subsection{General problem formulation}\label{sec:problem-formulate}
\begin{figure}
  \centering
  \includegraphics[width=0.45\textwidth]{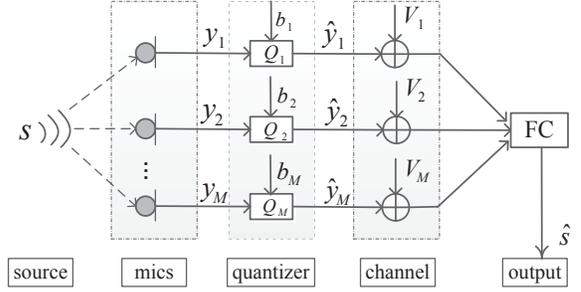}\\
  \caption{A typical communication model in WASNs.}
  \label{fig:commu_model}
\end{figure}
Fig.~\ref{fig:commu_model} shows a typical communication model in WASNs, which is considered in this work. The microphone recordings are quantized with specified bit rates and then transmitted to an FC through noisy communication channels. The FC conducts noise reduction and outputs the estimated target signal(s).
In this work, we are interested in minimizing the wireless transmission costs by allocating bit rates to microphones to achieve a prescribed noise reduction performance.
Our initial goal can be formulated in terms of the following optimization problem:
\begin{equation}\tag{P1}\label{eq:origin_opti}
\begin{aligned}
\min_{\mathbf{w}, \mathbf{b}} \ \ &\sum_{k=1}^{M}  d_k^2 V_k (4^{b_k}-1)\\
{\rm s.t.} \ \ & \mathbf{w}^H \mathbf{R}_{\mathbf{n+q}} \mathbf{w} \leq \frac{\beta}{\alpha}\\
& \mathbf{\Lambda}^H\mathbf{w}=\mathbf{f},\\
& b_k\in \mathbb{Z}_+, \quad b_k \leq b_0,\forall k,
\end{aligned}
\end{equation}
where $\beta$ denotes the minimum output noise power that can be achieved when all sensors are employed, $\alpha \in (0,1]$ is to control a certain expected performance, $\mathbb{Z}_+$ denotes a non-negative integer set, and $b_0$ the maximum rate per sample of each microphone signal.
Note that (\ref{eq:origin_opti}) is a general form for the rate-distributed spatial filtering based noise reduction problem. Also, $\beta/\alpha$ does not depend on the rate allocation strategy or statistics of the whole sensor network, because $\beta/\alpha$ is just a number that can be assigned by users, e.g., 40 dB, to indicate an expected performance.
By solving (\ref{eq:origin_opti}), we can determine the optimal rate distribution that each microphone can utilize to quantize its recordings, such that the noise reduction system achieves a desired performance with minimum energy usage.
One simple method to solve (\ref{eq:origin_opti}) is exhaustive search, i.e., evaluating the performance for all $(b_0+1)^M$ choices for the rate distribution, but evidently this is intractable unless $b_0$ or $M$ is very small.
Next, we will find an efficient solver for (\ref{eq:origin_opti}).

\subsection{Solver for rate-distributed LCMV beamforming}
In this section, we will reformulate (\ref{eq:origin_opti}) in the context of LCMV beamforming.
Considering the utilization of an LCMV beamformer for noise reduction,  the second constraint in (\ref{eq:origin_opti}) is automatically satisfied.
Substituting the solution of the LCMV beamformer from (\ref{eq:lcmv_formula}) into (\ref{eq:origin_opti}), we get the following simplified optimization problem:
\begin{equation}\label{eq:origin_opti_LCMV_simple}\tag{P2}
\begin{aligned}
\min_{\mathbf{b}} \ \ &\sum_{k=1}^{M}  d_k^2 V_k (4^{b_k}-1)\\
{\rm s.t.} \ \ &  \mathbf{f}^H \left(\mathbf{\Lambda}^H \mathbf{R}_{\mathbf{n+q}}^{-1}\mathbf{\Lambda}\right)^{-1} \mathbf{f} \leq \frac{\beta}{\alpha}\\
& b_k\in \mathbb{Z}_+, \quad b_k \leq b_0,\forall k,
\end{aligned}
\end{equation}
where the bit rates $\mathbf{b}$ are implicit in the output noise power $\mathbf{f}^H \left(\mathbf{\Lambda}^H \mathbf{R}_{\mathbf{n+q}}^{-1}\mathbf{\Lambda}\right)^{-1} \mathbf{f}$, which is clearly non-convex and non-linear in terms of $\mathbf{b}$. In what follows, we will explicitly express $\mathbf{f}^H \left(\mathbf{\Lambda}^H \mathbf{R}_{\mathbf{n+q}}^{-1}\mathbf{\Lambda}\right)^{-1} \mathbf{f}$ in $\mathbf{b}$ and reformulate (\ref{eq:origin_opti_LCMV_simple}) by semi-definite relaxation.

First of all, the first inequality constraint in (\ref{eq:origin_opti_LCMV_simple}) is equivalent to the following two new constraints by introducing a new symmetric positive semi-definite (PSD) matrix $\mathbf{Z}\in \mathbb{S}_{+}^{\mathcal{U}}$ with $\mathbb{S}_+$ denoting a set for symmetric PSD matrices, i.e.,
\begin{equation}\label{eq:multi-source-constrint2}
   \mathbf{\Lambda}^H\mathbf{R}_{\mathbf{n+q}}^{-1}\mathbf{\Lambda} = \mathbf{Z},
\end{equation}
\begin{equation}\label{eq:multi-source-constrint1}
   \mathbf{f}^H\mathbf{Z}^{-1}\mathbf{f} \leq  \frac{\beta}{\alpha}.
\end{equation}
The inequality (\ref{eq:multi-source-constrint1}) can be rewritten as a linear matrix inequality (LMI) using the Schur complement~\cite[p.650]{boyd2004convex}, i.e.,
\begin{equation}
\begin{bmatrix} \mathbf{Z}  & \mathbf{f} \\ \mathbf{f}^H   & \frac{\beta}{\alpha} \end{bmatrix} \succeq \mathbf{O}_{{\mathcal{U}}+1}.
\end{equation}
However, the equality constraint in (\ref{eq:multi-source-constrint2}) is clearly non-convex in terms of the unknowns $\mathbf{b}$. We therefore relax it to
\begin{equation}\label{eq:multi-source-constrint3}
   \mathbf{\Lambda}^H\mathbf{R}_{\mathbf{n+q}}^{-1}\mathbf{\Lambda} \succeq \mathbf{Z},
\end{equation}
since (\ref{eq:multi-source-constrint1}) and (\ref{eq:multi-source-constrint3}) are sufficient conditions to obtain the original constraint in (\ref{eq:origin_opti_LCMV_simple}), and we use $\succeq$ in (\ref{eq:multi-source-constrint3}) for convex relaxation.

Then, in order to linearize (\ref{eq:multi-source-constrint3}) in $\mathbf{b}$, we calculate $\mathbf{R}_{\mathbf{n+q}}^{-1}$ as
\begin{equation}\label{eq:inverseRnq}
\begin{aligned}[b]
\mathbf{R}_{\mathbf{n}+\mathbf{q}}^{-1} &= \left(\mathbf{R}_{\mathbf{nn}}+\mathbf{R_{qq}}\right)^{-1}\\
&= \mathbf{R}_{\mathbf{nn}}^{-1} - \mathbf{R}_{\mathbf{nn}}^{-1}\left(\mathbf{R}_{\mathbf{nn}}^{-1}+\mathbf{R}_{\mathbf{qq}}^{-1}  \right)^{-1}\mathbf{R}_{\mathbf{nn}}^{-1},
\end{aligned}
\end{equation}
where the second equality is derived from the matrix inversion lemma~\cite[p.18]{petersen2008matrix}
\begin{equation}\nonumber
\begin{aligned}
&\left(\mathbf{A}+\mathbf{CBC}^T \right)^{-1} = \\
&\qquad \mathbf{A}^{-1}-\mathbf{A}^{-1}\mathbf{C} \left(\mathbf{B}^{-1}+\mathbf{C}^T\mathbf{A}^{-1}\mathbf{C} \right)^{-1}\mathbf{C}^T\mathbf{A}^{-1}.
\end{aligned}
\end{equation}
Substitution of the expression for $\mathbf{R}_{\mathbf{n+q}}^{-1}$ from (\ref{eq:inverseRnq}) into (\ref{eq:multi-source-constrint3}), we obtain
\begin{equation}
   \mathbf{\Lambda}^H\mathbf{R}_{\mathbf{nn}}^{-1}\mathbf{\Lambda}  - \mathbf{Z} \succeq
   \mathbf{\Lambda}^H\mathbf{R}_{\mathbf{nn}}^{-1}\left(\mathbf{R}_{\mathbf{nn}}^{-1}+\mathbf{R}_{\mathbf{qq}}^{-1} \right)^{-1}\mathbf{R}_{\mathbf{nn}}^{-1}\mathbf{\Lambda}.
\end{equation}
Using the Schur complement, we obtain the following LMI\footnote{Note that (\ref{eq:constr2_opti_LCMV}) is not an LMI essentially, because it is not linear in the unknown parameters $\mathbf{b}$. Here, we call it LMI for convenience, since it looks like an LMI and is linear in $4^{b_k},\forall k$. }
\begin{equation}\label{eq:constr2_opti_LCMV}
\begin{bmatrix} \mathbf{R}_{\mathbf{nn}}^{-1}+\mathbf{R}_{\mathbf{qq}}^{-1} & \mathbf{R}_{\mathbf{nn}}^{-1}\mathbf{\Lambda}\\ \mathbf{\Lambda}^H\mathbf{R}_{\mathbf{nn}}^{-1}   & \mathbf{\Lambda}^H\mathbf{R}_{\mathbf{nn}}^{-1}\mathbf{\Lambda}  - \mathbf{Z} \end{bmatrix} \succeq \mathbf{O}_{M+\mathcal{U}},
\end{equation}
where $\mathbf{R}_{\mathbf{qq}}^{-1}$ can be computed from (\ref{eq:corr_matrix_quanti}) as
\begin{equation}
\begin{aligned}[b]
\mathbf{R}_{\mathbf{qq}}^{-1} =& 12\times {\rm diag}\left( \left[\frac{4^{b_1}}{\mathcal{A}_1^2}, \frac{4^{b_2}}{\mathcal{A}_2^2},...,\frac{4^{b_M}}{\mathcal{A}_M^2} \right] \right).
\end{aligned}
\end{equation}
For notational convenience, we define a constant vector $\mathbf{e}=\left[\frac{12}{\mathcal{A}_1^2},\cdots,\frac{12}{\mathcal{A}_M^2} \right]$. Further, we introduce a variable change  $t_k=4^{b_k} \in \mathbb{Z}_+,\forall k$, such that $\mathbf{R}_{\mathbf{qq}}^{-1}= {\rm diag}\left(\mathbf{e} \odot \mathbf{t} \right)$ and (\ref{eq:constr2_opti_LCMV}) are both linear in $\mathbf{t}$. In order to convexify the integer constraint $b_k\in \mathbb{Z}_+,\forall k$, we relax it to $b_k\in \mathbb{R}_+$, i.e., $t_k \in \mathbb{R}_+,\forall k$. Altogether, we arrive at
\begin{subequations}\label{eq:optimi_power_multi-source2}
\vspace{-0.05cm}
\begin{equation}
\min_{\mathbf{t}, \mathbf{Z}} \ \  \sum_{k=1}^{M}  d_k^2 V_k (t_k-1) \qquad \qquad \qquad \qquad \qquad \qquad
\tag{\ref{eq:optimi_power_multi-source2}}
\end{equation}
\vspace{-0.35cm}
\begin{align}
{\rm s.t. }\ \ &\begin{bmatrix} \mathbf{Z}  & \mathbf{f} \\ \mathbf{f}^H   & \frac{\beta}{\alpha} \end{bmatrix} \succeq \mathbf{O}_{{\mathcal{U}}+1},\label{eq:optimi_power_multi-source2_constraint1}\\
 & \begin{bmatrix} \mathbf{R}_{\mathbf{nn}}^{-1}+\mathbf{R}_{\mathbf{qq}}^{-1} & \mathbf{R}_{\mathbf{nn}}^{-1}\mathbf{\Lambda}\\ \mathbf{\Lambda}^H\mathbf{R}_{\mathbf{nn}}^{-1}   & \mathbf{\Lambda}^H\mathbf{R}_{\mathbf{nn}}^{-1}\mathbf{\Lambda}  - \mathbf{Z} \end{bmatrix} \succeq \mathbf{O}_{M+\mathcal{U}},\label{eq:optimi_power_multi-source2_constraint2}\\
  & 0\leq t_k \leq 4^{b_0}, \ \ \forall k, \label{eq:optimi_power_multi-source2_constraint3}
\end{align}
\end{subequations}
which is a standard semi-definite programming problem~\cite[p.128]{boyd2004convex} and can be solved efficiently in polynomial time using interior-point methods or solvers,  like CVX~\cite{grant2008cvx} or SeDuMi~\cite{sturm1999using}. The computational complexity for solving (\ref{eq:optimi_power_multi-source2}) is of the order of $\mathcal{O}((M+\mathcal{U})^3)$.

After (\ref{eq:optimi_power_multi-source2}) is solved, the allocated bit rates can be resolved by $b_k=\log_4 t_k,\forall k$ which are continuous values.
\subsection{Randomized rounding}\label{sec:random_rounding}
The solution provided by the semi-definite program in (\ref{eq:optimi_power_multi-source2}) consists of continuous values. A straightforward and often used technique to resolve the integer bit rates is by simply rounding, in which the integer estimates are given by ${\rm round} \left(b_k \right), \forall k$ where the ${\rm round}(\cdot)$ operator rounds its arguments towards the nearest integer. However, there is no guarantee that the integer solution obtained by this rounding technique always satisfies the performance constraint. Hence, we utilize a variant rounding technique, i.e., randomized rounding~\cite{chepuri2015sparsity}, to the estimates obtained from (\ref{eq:optimi_power_multi-source2}). Alternatively, we can use ${\rm ceil} \left(b_k \right), \forall k$ where the ${\rm ceil}(\cdot)$ operator rounds its arguments towards the nearest upper integer. However,  this is suboptimal compared to the randomized rounding technique due to more unnecessary energy usage.

\subsection{Representation of rate-distributed LCMV beamforming}\label{sec:represent_rateLCMV}
In this subsection, we will represent the rate-distributed LCMV beamforming in (\ref{eq:optimi_power_multi-source2}) from the perspective of Boolean optimization. This representation turns out to be very useful when comparing the rate-distributed LCMV beamforming framework to the LCMV beamforming based sensor selection framework. Setting $p_k=t_k/4^{b_0},\forall k$ in (\ref{eq:optimi_power_multi-source2}), we obtain the following equivalent form
\begin{subequations}\label{eq:optimi_power_multi-source3}
\begin{equation}
\min_{\mathbf{p},\mathbf{Z}} \ \ \frac{1}{4^{b_0}}\sum_{k=1}^{M} p_k V_k d_k^2 - {\rm const.} \qquad \qquad \qquad \qquad
\tag{\ref{eq:optimi_power_multi-source3}}
\end{equation}
\begin{align}
{\rm s.t.} \ \ &\begin{bmatrix} \mathbf{Z}  & \mathbf{f} \\ \mathbf{f}^H   & \frac{\beta}{\alpha} \end{bmatrix} \succeq \mathbf{O}_{{\mathcal{U}}+1},\label{eq:optimi_power_multi-source3_constraint1}\\
 & \begin{bmatrix}\mathbf{R}_{\mathbf{nn}}^{-1}+\mathbf{R}_{\mathbf{qq}}^{-1} & \mathbf{R}_{\mathbf{nn}}^{-1} \mathbf{\Lambda} \\ \mathbf{\Lambda}^H \mathbf{R}_{\mathbf{nn}}^{-1} & \mathbf{\Lambda}^H\mathbf{R}_{\mathbf{nn}}^{-1}\mathbf{\Lambda} -\frac{\alpha}{\beta} \end{bmatrix}  \succeq \mathbf{0}_{M+\mathcal{U}},\label{eq:optimi_power_multi-source3_constraint2}\\
 & 0 \leq p_k \leq 1,\forall k, \label{eq:optimi_power_multi-source3_constraint3}
\end{align}
\end{subequations}
where $\mathbf{R}_{\mathbf{qq}}^{-1}= 4^{b_0}{\rm diag}\left(\mathbf{e} \odot \mathbf{p} \right)$.  Note that for (\ref{eq:optimi_power_multi-source3}), minimizing $\frac{1}{4^{b_0}}\sum_{k=1}^{M} p_k V_k d_k^2 - {\rm const}$ is equivalent to minimizing $\sum_{k=1}^{M} p_k V_k d_k^2$. Given the solution of (\ref{eq:optimi_power_multi-source3}), the rates to be allocated can be resolved by $b_k = \log_4 p_k + b_0, \forall k$ and the randomized rounding technique in Sec.~\ref{sec:random_rounding}.
\begin{remark}
From the perspective of optimization, (\ref{eq:optimi_power_multi-source2}) and (\ref{eq:optimi_power_multi-source3}) are equivalent, i.e., both are semi-definite programming problems with the same computational complexity and can provide the optimal rate distribution. However, apart from the function of rate allocation, (\ref{eq:optimi_power_multi-source3}) gives an insight to sensor selection, because its unknowns $\mathbf{p}$ are Boolean variables which can indicate whether a sensor is selected or not. In other words, if we are interested in sparsity-aware networks instead of energy-aware ones, (\ref{eq:optimi_power_multi-source3}) can be employed to select the best microphone subset.
\end{remark}
Based on the representation of rate-distributed LCMV beamforming in (\ref{eq:optimi_power_multi-source3}), we will find the relation between  rate allocation and sensor selection  in the next section.

\section{Relation to microphone subset selection}\label{sec:rateLCMV_vs_SenSel}
In this section, we will show the relation between rate allocation and sensor selection. To do so, we extend the sensor selection based MVDR beamformer from~\cite{zhang2017microphone} to the LCMV beamforming framework. We first find that sensor selection is a special case of the rate allocation problem. Then, we propose a bisection algorithm that can be used to obtain the sensor selection results as in~\cite{zhang2017microphone} based on the rate allocation method.

\subsection{Model-driven LCMV beamforming}\label{sec:MD-LCMV}
In~\cite{zhang2017microphone}, we considered the problem of microphone subset selection based noise reduction in the context of MVDR beamforming. We minimized the transmission costs by constraining to a desired noise reduction performance. The transmission cost was  related to the distance between each microphone and the FC.  In the case the number of constraints in (\ref{eq:lcmv_opti}) is reduced to a single constraint preserving a single target, the LCVM beamformer reduces to a special case, i.e., the MVDR beamformer. Hence, mathematically, the original sensor selection problem in~\cite{zhang2017microphone} can be extended by adding more linear constraints to obtain the following optimization problem
\begin{equation}\label{eq:SeSel1}
\begin{aligned}
  \min_{\mathbf{w_p},\mathbf{p}} \ \  &\sum_{k=1}^M p_k d_k^2 \\
  {\rm s. t.} \ \ &\mathbf{w}^H_{\mathbf{p}}\mathbf{R_{n+q,p}}\mathbf{w_p} \leq \frac{\beta}{\alpha},\\
  &\mathbf{\Lambda}^H_{\mathbf{p}}\mathbf{w_p}=\mathbf{f},
\end{aligned}
\end{equation}
where $\mathbf{p}=[p_1,\cdots,p_M]^T\in \{0,1 \}^M$ are selection variables to indicate whether a sensor is selected or not, $\mathbf{w}_{\mathbf{p}}$  denotes the coefficients of the LCMV beamformer corresponding to the selected sensors, $\mathbf{\Lambda}_{\mathbf{p}}$ is a submatrix of $\mathbf{\Lambda}$ which was defined in (\ref{eq:lcmv_opti}), and other parameters are defined similarly as in (\ref{eq:origin_opti}). Suppose that for the microphone subset selection problem, all the candidate sensors use the maximum rates, i.e., $b_0$ bits per sample, to communicate with the FC, such that $\mathbf{R_{n+q}} = \mathbf{R_{nn}}+\mathbf{R_{qq}}$ and $\mathbf{R_{qq}}=\frac{1}{12}\times {\rm diag}\left( \left[\frac{\mathcal{A}_1^2}{4^{b_0}}, \frac{\mathcal{A}_2^2}{4^{b_0}},...,\frac{\mathcal{A}_M^2}{4^{b_0}} \right] \right)$. The problem (\ref{eq:SeSel1}) is called model-driven LCMV beamforming, because it is based on the statistical knowledge $\mathbf{R_{n+q}}$.

We will show that the optimization problem in (\ref{eq:SeSel1}) can be solved by considering (\ref{eq:optimi_power_multi-source3}).
Let ${\rm diag}(\mathbf{p})$ be a diagonal matrix whose diagonal entries are given by $\mathbf{p}$, such that $\mathbf{\Phi}_{\mathbf{p}}\in \{0,1\}^{K\times M}$ is a submatrix of ${\rm diag}(\mathbf{p})$ after all-zero rows (corresponding to the unselected sensors) have been removed. As a result, we can easily get the following relationships
\begin{equation}
\begin{aligned}
  \mathbf{\Phi}_{\mathbf{p}}\mathbf{\Phi}_{\mathbf{p}}^T  = \mathbf{I}_K, \ \
  \mathbf{\Phi}_{\mathbf{p}}^T\mathbf{\Phi}_{\mathbf{p}}  = {\rm diag}(\mathbf{p}).
\end{aligned}
\end{equation}
Therefore, applying the selection model to the classical LCMV beamformer in (\ref{eq:lcmv_formula}), the best linear unbiased estimator for a subset of $K$ microphones determined by $\mathbf{p}$ will be
\begin{equation}\label{eq:LCMV-ss}
  \hat{\mathbf{w}}_{\mathbf{p}}=\mathbf{R}_{\mathbf{n+q,p}}^{-1}\mathbf{\Lambda}_{\mathbf{p}}\left(\mathbf{\Lambda}^H_{\mathbf{p}} \mathbf{R}_{\mathbf{n+q,p}}^{-1}\mathbf{\Lambda}_{\mathbf{p}}\right)^{-1}\mathbf{f} ,
\end{equation}
where $ \mathbf{R}_{\mathbf{n+q,p}} = \mathbf{\Phi}_{\mathbf{p}}\mathbf{R}_{\mathbf{n+q}}\mathbf{\Phi}_{\mathbf{p}}^T$
represents the total noise correlation matrix of the selected sensors after the rows and columns of $\mathbf{R}_{\mathbf{n+q}}$ corresponding to the unselected sensors have been removed, i.e., $ \mathbf{R}_{\mathbf{n+q,p}}$ is a submatrix of $\mathbf{R}_{\mathbf{n+q}}$.

Applying the result in (\ref{eq:LCMV-ss}) to (\ref{eq:SeSel1}) yields a simplified optimization problem based on the LCMV beamformer as
\begin{equation}\label{eq:SeSel2}
\begin{aligned}
  \min_{\mathbf{p}} \ \  &\sum_{i=1}^M p_i d_i^2 \\
  {\rm s. t.} \ \ &\mathbf{w}^H_{\mathbf{p}}\mathbf{R_{n+q,p}}\mathbf{w_p} \leq \frac{\beta}{\alpha},
\end{aligned}
\end{equation}
where similar to (\ref{eq:out_npower_lcmv}) the output noise power is given by
\begin{equation}
\mathbf{w}^H_{\mathbf{p}}\mathbf{R_{n+q,p}}\mathbf{w_p} = \mathbf{f}^H\left( \mathbf{\Lambda}_{\mathbf{p}}^H\mathbf{R}_{\mathbf{n+q,p}}^{-1}\mathbf{\Lambda}_{\mathbf{p}}\right)^{-1}\mathbf{f}.
\end{equation}
By introducing a symmetric PSD matrix $\mathbf{Z}\in\mathbb{S}_+^{\mathcal{U}}$, we can rewrite the constraint in (\ref{eq:SeSel2}) into two new constraints in a similar way as in the previous section, i.e.,
\begin{equation}\label{eq:SeSel-constrint2}
   \mathbf{\Lambda}^H\mathbf{R}_{\mathbf{n+q}}^{-1}\mathbf{\Lambda} = \mathbf{Z},
\end{equation}
\begin{equation}\label{eq:SeSel-constrint1}
   \mathbf{f}^H\mathbf{Z}^{-1}\mathbf{f} \leq  \frac{\beta}{\alpha}.
\end{equation}
The inequality in (\ref{eq:SeSel-constrint1}) can be rewritten as an LMI using the Schur complement, which is identical to (\ref{eq:optimi_power_multi-source3_constraint1}).
Also, similar to Sec.~\ref{sec:rateLCMV}, we relax the equality constraint in (\ref{eq:SeSel-constrint2}) to
\begin{equation}\label{eq:SeSel-constrint3}
   \mathbf{\Lambda}_{\mathbf{p}}^H\mathbf{R}_{\mathbf{n+q,p}}^{-1}\mathbf{\Lambda}_{\mathbf{p}} \succeq \mathbf{Z},
\end{equation}
due to the non-convexity. The left side of (\ref{eq:SeSel-constrint3}) can be calculated as
\begin{equation}\label{eq:SeSel-constraint4}
\begin{aligned}[t]
&\mathbf{\Lambda}_{\mathbf{p}}^H\mathbf{R}_{\mathbf{n+q,p}}^{-1}\mathbf{\Lambda}_{\mathbf{p}} \overset{\text{(a)}}{=} \mathbf{\Lambda}^H\mathbf{\Phi}_{\mathbf{p}}^T\mathbf{R}_{\mathbf{n+q,p}}^{-1}\mathbf{\Phi}_{\mathbf{p}}\mathbf{\Lambda} \\
&\overset{\text{(b)}}{=} \mathbf{\Lambda}^H\mathbf{\Phi}_{\mathbf{p}}^T\left(\mathbf{\Phi}_{\mathbf{p}} \mathbf{R}_{\mathbf{n+q}}\mathbf{\Phi}_{\mathbf{p}}^T  \right)^{-1}\mathbf{\Phi}_{\mathbf{p}}\mathbf{\Lambda}\\
&\overset{\text{(c)}}{=} \mathbf{\Lambda}^H\mathbf{\Phi}_{\mathbf{p}}^T\left(\mathbf{\Phi}_{\mathbf{p}} \mathbf{R}_{\mathbf{nn}}\mathbf{\Phi}_{\mathbf{p}}^T   +\underbrace{\mathbf{\Phi}_{\mathbf{p}} \mathbf{R_{qq}}\mathbf{\Phi}_{\mathbf{p}}^T}_{\mathbf{Q}}  \right)^{-1}\mathbf{\Phi}_{\mathbf{p}}\mathbf{\Lambda}\\
&\overset{\text{(d)}}{=} \mathbf{\Lambda}^H\left[\mathbf{R}_{\mathbf{nn}}^{-1}-\mathbf{R}_{\mathbf{nn}}^{-1}\left(\mathbf{R}_{\mathbf{nn}}^{-1}+ \mathbf{\Phi}_{\mathbf{p}}^T \mathbf{Q}^{-1}\mathbf{\Phi}_{\mathbf{p}} \right)^{-1}\mathbf{R}_{\mathbf{nn}}^{-1}\right]\mathbf{\Lambda}\\
&\overset{\text{(e)}}{=}\mathbf{\Lambda}^H\mathbf{R}_{\mathbf{nn}}^{-1}\mathbf{\Lambda}-\mathbf{\Lambda}^H\mathbf{R}_{\mathbf{nn}}^{-1}\left(\mathbf{R}_{\mathbf{nn}}^{-1}+ 4^{b_0}{\rm diag}(\mathbf{p} \odot \mathbf{e}) \right)^{-1}\mathbf{R}_{\mathbf{nn}}^{-1}\mathbf{\Lambda},
\end{aligned}
\end{equation}
where (c) constructs $\mathbf{\Phi}_{\mathbf{p}} \mathbf{R_{qq}}\mathbf{\Phi}_{\mathbf{p}}^T$ as a new diagonal matrix $\mathbf{Q}\in \mathbb{R}^{K\times K}$ whose diagonal entries correspond to the selected sensors, (d) is derived based on the matrix inversion lemma~\cite[p.18]{petersen2008matrix}\footnote{Based on the Woodbury identity $\left(\mathbf{A}+\mathbf{CBC}^T\right)^{-1} = \mathbf{A}^{-1}-\mathbf{A}^{-1}\mathbf{C} \left(\mathbf{B}^{-1}+\mathbf{C}^T\mathbf{A}^{-1}\mathbf{C} \right)^{-1}\mathbf{C}^T\mathbf{A}^{-1}$, we can see that $\mathbf{C} \left(\mathbf{B}^{-1}+\mathbf{C}^T\mathbf{A}^{-1}\mathbf{C} \right)^{-1}\mathbf{C}^T = \mathbf{A}-\mathbf{A}\left(\mathbf{A}+\mathbf{CBC}^T\right)^{-1}\mathbf{A}$. Taking $\mathbf{A}=\mathbf{R}_{\mathbf{nn}}^{-1},\mathbf{B}=\mathbf{Q}^{-1}$ and $\mathbf{C}=\mathbf{\Phi}_{\mathbf{p}}^T$ and applying the Woodbury identity to the right side of the third equality in (\ref{eq:SeSel-constraint4}), we can obtain the fourth equality. }, and (e) holds for $\mathbf{p}$ are Boolean variables.

Substitution of  (\ref{eq:SeSel-constraint4}) into (\ref{eq:SeSel-constrint3}) and using the Schur complement, we can obtain an LMI which will be identical to (\ref{eq:optimi_power_multi-source3_constraint2}). Altogether, we then reformulate the sensor selection problem for the LCMV beamforming as the following semi-definite program:
\vspace{-0.2cm}
\begin{subequations}\label{eq:SeSel3}
\begin{equation}
\min_{\mathbf{p}} \ \ \sum_{k=1}^{M} p_k  d_k^2  \qquad \qquad \qquad \qquad \qquad \qquad \qquad
\tag{\ref{eq:SeSel3}}
\end{equation}
\vspace{-0.3cm}
\begin{align}
{\rm s.t.} \ \ &\begin{bmatrix} \mathbf{Z}  & \mathbf{f} \\ \mathbf{f}^H   & \frac{\beta}{\alpha} \end{bmatrix} \succeq \mathbf{O}_{{\mathcal{U}}+1},\label{eq:SeSel3_constraint1}\\
 & \begin{bmatrix}\mathbf{R}_{\mathbf{nn}}^{-1}+\mathbf{R}_{\mathbf{qq}}^{-1} & \mathbf{R}_{\mathbf{nn}}^{-1} \mathbf{\Lambda} \\ \mathbf{\Lambda}^H \mathbf{R}_{\mathbf{nn}}^{-1} & \mathbf{\Lambda}^H\mathbf{R}_{\mathbf{nn}}^{-1}\mathbf{\Lambda} -\frac{\alpha}{\beta} \end{bmatrix}  \succeq \mathbf{0}_{M+\mathcal{U}},\label{eq:SeSel3_constraint2}\\
 & 0 \leq p_k \leq 1,\forall k, \label{eq:SeSel3_constraint3}
\end{align}
\end{subequations}
where the Boolean variables $p_k, \forall k$ have already been relaxed by continuous surrogates. As a consequence, we see that the sensor selection problem in (\ref{eq:SeSel3}) is equivalent to the rate allocation  problem in (\ref{eq:optimi_power_multi-source3}) when all the communication channels have the same noise power, e.g., $V_k=1,\forall k$. Based on this observation, it can be concluded that the sensor selection problem can be solved by the rate allocation algorithm. In other words, the proposed rate allocation approach is a generalization of the sensor selection method in~\cite{zhang2017microphone}.
\vspace{-0.2cm}
\subsection{Threshold determination by bisection algorithm}\label{sec:bisection_algo}
In Sec.~\ref{sec:MD-LCMV}, we have shown the relationship between the rate allocation problem and sensor selection, i.e., the former is a generalization of the latter problem, from a theoretical perspective. From this, we know that the best subset of microphones can be identified by the solution of rate distribution. Now, the essential question remaining is how to determine the selected sensors as in~\cite{zhang2017microphone}, based on the rate distribution presented in the current work.  Here, we propose a bisection algorithm for threshold determination.

In detail, given the rate distribution $b_k, \forall k$ which is the solution of the problem (\ref{eq:optimi_power_multi-source2})  and the maximum rate $b_0$, first we set the threshold $T=\frac{b_0}{2}$, such that we choose a subset of sensors, say $\mathcal{S}$, whose rate is larger than $T$, that is, $\mathcal{S}=\{k|b_k \ge T\}$. If the performance using the sensors contained in the set $\mathcal{S}$, say $\tau$, is larger than $\frac{\beta}{\alpha}$, we decrease $T$ and update $\mathcal{S}$; if $\tau < \frac{\beta}{\alpha}$, we will increase $T$. This procedure continues until $\frac{\beta}{\alpha}-\tau \leq \epsilon$ where $\epsilon$ is a predefined very small positive number.
%
Furthermore, the best subset of microphones can also be found by solving the optimization problem in (\ref{eq:optimi_power_multi-source3}), while we need to apply the randomized rounding technique to resolve the Boolean variables $\mathbf{p}$. Usually, the bisection algorithm is much faster than randomized rounding.
\vspace{-0.2cm}
\section{Numerical results}\label{sec:exp_example}
In this section, we will show some numerical results for the proposed algorithm in terms of noise reduction in WASNs.
\begin{figure}
  \centering
  \vspace{-0.4cm}
  \includegraphics[width=0.45\textwidth]{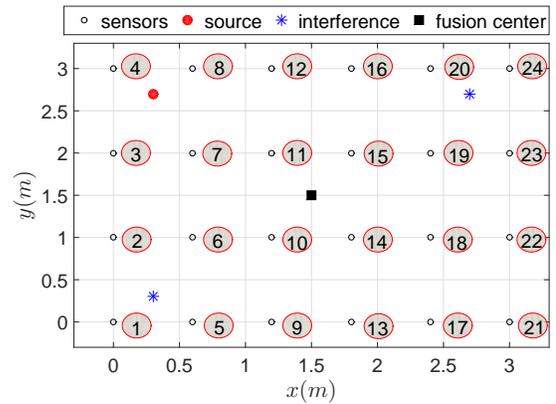}\\
  \caption{A typical wireless acoustic sensor network in a 2D scenario, where the indexes of microphones are labelled.}
  \vspace{-0.4cm}
  \label{fig:exp_setup}
\end{figure}

\subsection{Single target source}
Fig.~\ref{fig:exp_setup} shows the experimental setup employed in the simulations, where 24 candidate microphones are placed uniformly in a 2D room with dimensions $(3\times 3)$ m. The desired speech source (red solid circle) is located at $(0.3, 2.7)$ m. The FC (black solid square) is placed at the centre of the room. Two interfering sources (blue stars) are positioned at $(0.3,0.3)$ m and $(2.7,2.7)$ m, respectively. The target source signal is a 10 minute long concatenation of speech signals originating from the TIMIT database~\cite{garofolo1988darpa}. The interferences are stationary Gaussian speech shaped noise sources. The uncorrelated noise is modeled as microphone self noise at an SNR of 50 dB. All signals are sampled at 16 kHz. We use a square-root Hann window of 20 ms for framing with 50\% overlap. The acoustic transfer functions are generated using~\cite{habets2006room} with reverberation time $T_{60}=200$ ms. In order to focus on the rate-distributed spatial filtering issue, we assume that a perfect voice activity detector (VAD) is available in the sequel. For the noise correlation matrix $\mathbf{R_{nn}}$, it is estimated at the FC end using sufficient-long noise-only segments when each node communicates with the FC at the maximum rate $b_0$ or larger.

\begin{figure}
  \centering
  \includegraphics[width=0.45\textwidth]{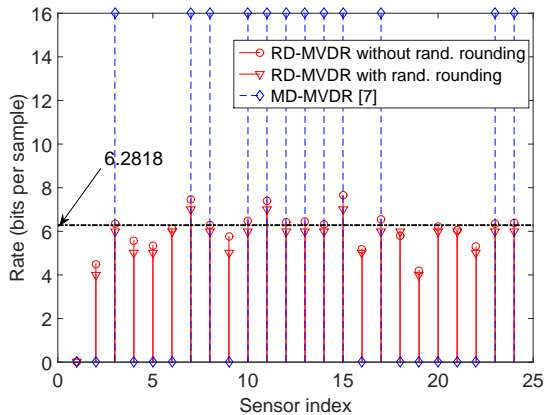}\\
  \vspace{-0.2cm}
  \caption{Example of  bit-rate allocation by the proposed approach (RD-MVDR) and microphone subset selection method (MD-MVDR). For the latter method, the selected sensors are allocated with $b_0$ bits, i.e., 16 bits per sample.}
  \vspace{-0.3cm}
  \label{fig:blind_rates-allo}
\end{figure}

An example of bit-rate allocation obtained by the rate-distributed LCMV beamforming and model-driven
sensor selection based MVDR beamforming (referred to as MD-MVDR in short)~\cite{zhang2017microphone} is shown in  Fig.~\ref{fig:blind_rates-allo} with $\alpha=0.8$. Since only one target source of interest exists, the optimization problem in (\ref{eq:optimi_power_multi-source2}) for the proposed method reduces to rate-distributed MVDR beamforming, which is referred to as RD-MVDR in short. From Fig.~\ref{fig:blind_rates-allo}, it is observed that in order to fulfill the same performance, the proposed RD-MVDR method activates more sensors than the MD-MVDR. The MD-MVDR has a smaller cardinality of the selected subset. However, each active sensor obtained by RD-MVDR is allocated with a much lower bit-rate per sample compared to the maximum rates, i.e., $b_0=16$ bits. Also, the sensors that are close to the target source and the FC are more likely to be allocated with higher bit-rates, because they have a higher SNR and less energy costs, respectively. More importantly, we find a threshold for the rate distribution of RD-MVDR, e.g., 6.2818 bits, using the bisection algorithm from Sec.~\ref{sec:bisection_algo}, and the active sensors whose rates are larger than this threshold are completely the same as the best subset obtained using the MD-MVDR algorithm. This phenomenon supports the conclusion that we have made in Sec.~\ref{sec:rateLCMV_vs_SenSel}, i.e., the best microphone subset selection problem can be resolved by the rate allocation algorithm. Hence, given the solution of rate distribution, to find out the best microphone subset is equivalent to determining a bit-rate threshold.

\begin{figure}
  \centering
  \includegraphics[width=0.45\textwidth]{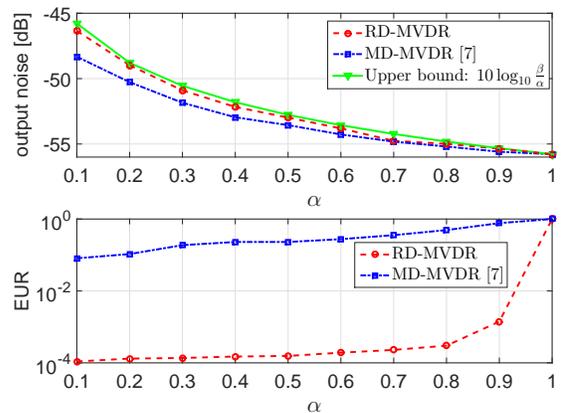}\\
  \vspace{-0.2cm}
  \caption{Output noise power and energy usage ratio (EUR) in terms of $\alpha$.  In the log-domain, the gap between the desired performance (i.e., $\beta/\alpha$) and the maximum performance when using all sensors (i.e., $\beta$) will be $-10\log_{10}\alpha$.}
\vspace{-0.3cm}
  \label{fig:result_WASN}
\end{figure}

In order to show the comparison of the proposed method in terms of noise reduction and energy usage, we also show the output noise power (in dB) and energy usage ratio (EUR) in terms of $\alpha$ in Fig.~\ref{fig:result_WASN}, where the indicator EUR is defined by
\[
{\rm EUR}_i = E_i/E_{\max}, \quad i\in \{\text{ RD-MVDR, MD-MVDR}\},
\]
where $E_i$ denotes the energy used by the RD-MVDR or MD-MVDR method, and $E_{\max}$ the maximum transmission energy when all the sensors are involved and communicate with the FC using $b_0$ bits. Clearly, the lower the EUR, the better the energy efficiency. In Fig.~\ref{fig:result_WASN}, we also compare to the desired maximum noise power, i.e., $10\log_{10}\frac{\beta}{\alpha}$. Note that $\beta$ denotes the output noise power when using all sensors. Although this is hard to calculate in practice, in the simulations it can be estimated by including all sensors and allocating each with $b_0$ bits. In practical applications, we just need to set a value for $10\log_{10}\frac{\beta}{\alpha}$, e.g., 40 dB, to constrain the desired performance. From Fig.~\ref{fig:result_WASN}, it follows that both RD-MVDR and MD-MVDR satisfy the performance requirement (i.e., below the upper bound $10\log_{10}\frac{\beta}{\alpha}$), while RD-MVDR is more efficient in the sense of energy usage, which is also explicit in the rate distribution in Fig.~\ref{fig:blind_rates-allo}.

\subsection{Multiple target sources}
\begin{figure}
  \centering
 \includegraphics[width=0.45\textwidth]{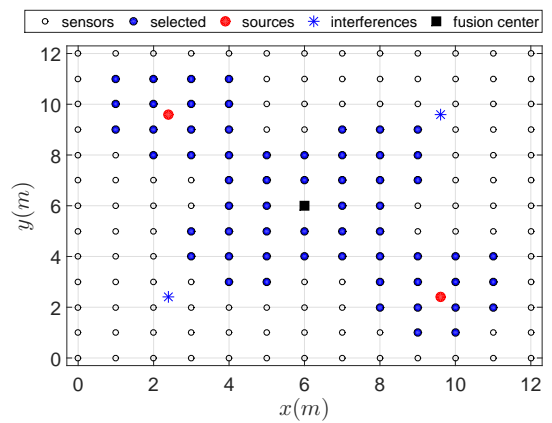}\\
 \vspace{-0.3cm}
  \caption{A larger-scale WASN, which consists of 169 microphone uniformly placed in a ($12\times 12$) m 2D room. The sensors are labelled from bottom to top and from left to right, which is similar to the labeling in Fig.~\ref{fig:exp_setup}. The selected microphones are obtained by solving (\ref{eq:SeSel3}) for $\alpha=0.8$.}
  \vspace{-0.3cm}
  \label{fig:SeSel_largeWASN}
\end{figure}

\begin{figure*}
  \centering
 \includegraphics[width=1.1\textwidth,height=0.44\textwidth]{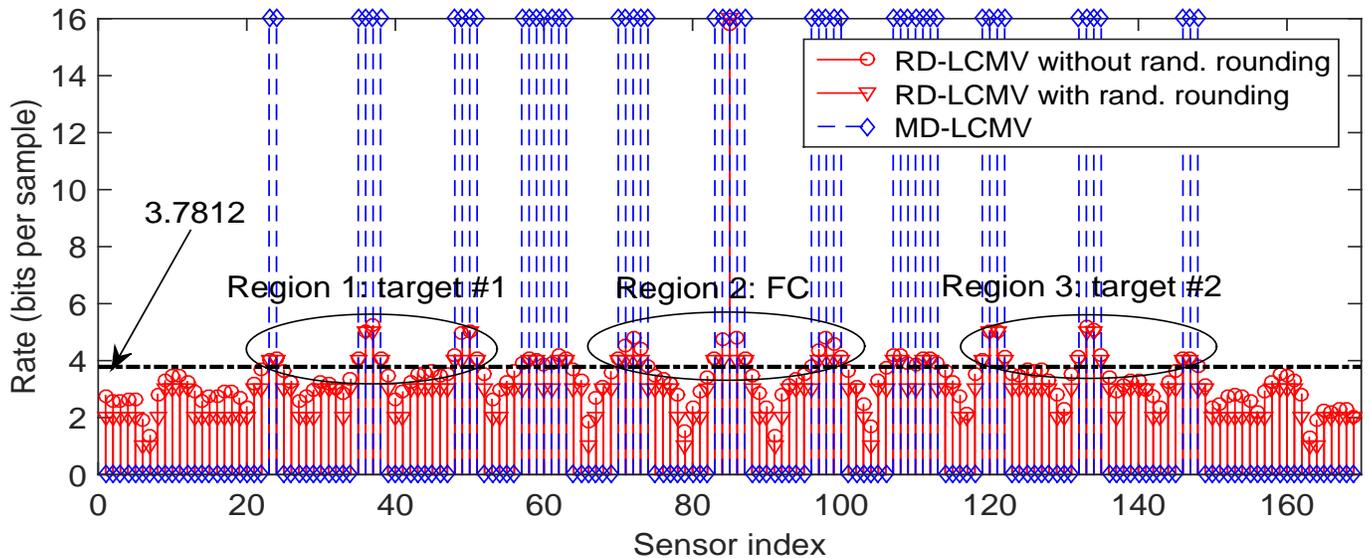}\\
  \caption{Rate distribution for the larger-scale WASN in Fig.~\ref{fig:SeSel_largeWASN} with $\alpha=0.8$. The MD-LCMV problem is solved by the bisection algorithm using the results of RD-LCMV. Clearly, the sensors within three regions that are close to the targets and the FC are allocated with higher rates.}
  \label{fig:rate_largeWASN}
\end{figure*}

In order to further investigate the noise reduction capability of the proposed algorithm for multiple target sources, we consider a larger-scale WASN as Fig.~\ref{fig:SeSel_largeWASN} shows, which consists of 169 microphones uniformly placed in a 2D room with dimensions ($12\times 12$) m. The FC is placed at the center of the room. Two target sources are located at $(2.4,9.6)$ m and $(9.6,2.4)$ m, respectively. Two interfering sources are located at $(2.4,2.4)$ m and $(9.6,9.6)$ m, respectively. Fig.~\ref{fig:rate_largeWASN} shows the rate distribution, where the proposed method (referred as RD-LCMV) is compared to the model-driven sensor selection method (referred as MD-LCMV in Sec.~\ref{sec:MD-LCMV}), which is solved by the bisection algorithm in Sec.~\ref{sec:bisection_algo}. Similar to Fig.~\ref{fig:blind_rates-allo}, the sensors that are close to the target sources and FC are allocated with higher rates. The 85th microphone node is allocated with the highest rate, e.g., 16 bits, because it is exactly located at the position of the FC. Also, it is shown that the best microphone subset by MD-LCMV can be determined by finding the optimal threshold for the solution of RD-LCMV (i.e., 3.7812 bits). Furthermore, we plot the sensor selection result that is obtained by solving (\ref{eq:SeSel3})  in Fig.~\ref{fig:SeSel_largeWASN}. Comparing the sensors selected by solving (\ref{eq:SeSel3}) as shown in Fig.~\ref{fig:SeSel_largeWASN}  to the sensors that are selected by applying the bisection algorithm to the solution of the RD-LCMV algorthm as shown in Fig.~\ref{fig:rate_largeWASN}, we see that both sets are completely identical. This also validates the relationship between sensor selection and the rate allocation problem.

To summarize, the rate allocation algorithms (RD-LCMV or RD-MVDR) activate more sensors than the sensor selection algorithms (MD-MVDR or MD-LCMV) in general, but each activated sensor is allocated with a much lower bit-rate. Hence, from the perspective of energy usage for data transmission, the rate allocation algorithms consume less energy.

\section{Conclusion}\label{sec:conclusion}
In this paper, we investigated the rate-distributed spatial filtering based noise reduction problem in energy-aware WASNs. A good strategy for bit-rate allocation can significantly save the energy costs, and meanwhile achieve a prescribed noise reduction performance as compared to a blindly uniform allocation for the best microphone subset obtained by the sensor selection approach. The problem was formulated by minimizing the total transmission costs subject to the constraint on a desired performance. In the context of LCMV beamforming, we formulated the problem as a semi-definite program (i.e., RD-LCMV). Further, we extended the model-driven sensor selection approach in~\cite{zhang2017microphone} for the LCMV beamforming (i.e., MD-LCMV). It was shown that the rate allocation problem is a generalization of sensor selection, e.g., the best subset of microphones can be chosen by determining the optimal threshold for the rates that are obtained by the RD-LCMV or RD-MVDR algorithm. In WASNs, based on numerical validation, we found that the microphones that are close to the source(s) and the FC are allocated with higher rates, because they are helpful for signal estimation and for reducing energy usage, respectively.

\end{document}